\documentclass[aps,twocolumn,showpacs]{revtex4}
\usepackage{dcolumn}
\usepackage{float}
\usepackage{graphicx}
\usepackage{amsmath}
\usepackage{verbatim}
\usepackage{amssymb}
\usepackage{psfrag}
\usepackage{color}
\usepackage{bm}
\usepackage{wrapfig}
\usepackage{subfigure}
\usepackage{makeidx}
\usepackage{bm}
\usepackage{epsf}
\usepackage{multirow}
\usepackage{mathrsfs}
\usepackage{braket}
\usepackage[colorlinks,linkcolor=blue,urlcolor=blue,citecolor=blue]{hyperref}

\DeclareMathOperator{\sech}{sech}
\begin{document}
\title{Soliton Thouless pumping engineered by inter-site nonlinearities}
\author{Tao Jiang$^{1}$}
\author{Li-Chen Zhao$^{1,2,3}$}\email{zhaolichen3@nwu.edu.cn}

\address{$^{1}$School of Physics, Northwest University, Xi'an, 710127, China}
\address{$^{2}$NSFC-SPTP Peng Huanwu Center for Fundamental Theory, Xi'an 710127, China}
\address{$^{3}$Shaanxi Key Laboratory for Theoretical Physics Frontiers, Xi'an 710127, China}

\begin{abstract}

We study soliton Thouless pumping in an extended diagonal Aubry-Andr\'e-Harper model with on-site nonlinearities and inter-site nonlinearities. We show that the inter-site nonlinearities can make solitons acquire anomalous transport distances far beyond  the ones predicted by the linear bands, and the quantized  displacements can be engineered well.  We  uncover that nonlinear instabilities require lower limits on sweeping rates for soliton pumping, challenging the common notion that slower modulation enables a more favorable realization of topological transport. The nonlinear interactions between solitons make multi-soliton pumping generally lack the robustness characteristic of Thouless pumping as linear systems. Our results provide many possibilities to engineer topological pumping by nonlinearities, and further make a step for applications of soliton pumping.

\end{abstract}

\date{\today}
\maketitle

\section{Introduction}
Soliton Thouless pumping is investigated both experimentally \cite{Jurgensen 2021,Jurgensen 2023} and theoretically \cite{Jurgensen 2022,Ye Fangwei 2022,Ye Fangwei2 2022,Wannier soliton,Gong Jiangbin 2023,Z. Liang 2024,NLin CNs 2025} in nonlinear systems, which greatly extends the applicable scope of topological pumping \cite{Thouless 1983,Thouless 1984,Niu Qian 2010}. Nonlinearities mainly localize the wave packets to be solitons, but do not bring any additional topology beyond the linear bands for the most of previously reported soliton pumping \cite{Jurgensen 2021,Jurgensen 2023, Jurgensen 2022,Ye Fangwei 2022,Ye Fangwei2 2022,Wannier soliton,Gong Jiangbin 2023,Z. Liang 2024,NLin CNs 2025}. Namely, the quantized transport distance can still be described by a linear single-band Chen number (CN)  or a combination of multi-band CNs. However, nonlinear geometric phases \cite{Niu 2005,Liu 2010} and some exotic topology-related phenomena induced by nonlinearities \cite{Non-ind1 2013,Non-ind2 2020,Non-ind3 2016,Non-ind4 2020,Non-ind5 2024,Non-ind6 2016}, suggest that nonlinearities could induce topological pumping beyond linear band topology. Very recently,  nonlinearity-induced soliton transport was indeed reported in some nonlinear systems \cite{Xu Yong1 2025,Xu Yong2 2025}. Considering that there are varied nonlinearities in real physical systems,  we would like to further look for exotic nonlinearities-induced topological pumping, and even try to engineer soliton pumping by setting proper nonlinearities.

Adiabaticity is usually an essential condition for achieving topological transport \cite{Thouless 1983,Thouless 1984,Niu Qian 2010}. Generally, it is believed that slower modulation enables a more favorable realization of topological transport. However, there could be some nonlinear instabilities during the soliton pumping processes. Even if the system is driven adiabatically, such instabilities can disrupt the topological transport of solitons \cite{Jurgensen 2022}. Fortunately,  the instabilities destructive effect takes a certain amount of time to accumulate before it can manifest itself, and the magnitude of instabilities depend on the soliton forms and many physical parameters \cite{book 2003, Solitons 2011}.  Therefore, there could be some competitions between instabilities and adiabaticity requirements. How to addressing them properly is one of the most crucial factors for applications of soliton pumping.

We consider a diagonal Aubry-Andr\'e-Harper (AAH) model \cite{AAH, Harper 1955} with taking both on-site nonlinearities (OSN) and inter-site nonlinearities (ISN), to investigate soliton Thouless transport.  When only OSN is present, the quantized transport distance is fully determined by the CN of linear bands, which agrees well with the related previous studies \cite{Jurgensen 2021,Jurgensen 2023, Jurgensen 2022,Ye Fangwei 2022,Ye Fangwei2 2022,Wannier soliton,Gong Jiangbin 2023,Z. Liang 2024,NLin CNs 2025}. Strikingly, when increasing strengthes of ISN, solitons can possess displacements of 2, 3, or even 4 unit cells  within a single pumping cycle, which far exceeding the CN$=+1$ of the corresponding linear band. Remarkably, the soliton with extra transport does not occupy multiple linear bands during its evolution, in contrast to the multiple pumps reported before \cite{Ye Fangwei 2022}. These additional quantized displacements arise  entirely from ISN, and can be engineered well by varying the ISN strengthes. We further analyze the soliton instabilities during the transport process, and find that instabilities bring clear lower limits on sweeping rates. While adiabaticity must be maintained, faster driving above the lower limits may allow the soliton to inhibit instabilities development, and facilitate the completion of topological pumping. Finally, we investigate multi-soliton transport and find that the nonlinear interactions between solitons make the pumping process generally less robust than Thouless pumping in linear systems,  which could provide alternative ways to manipulating pumping dynamics.

The paper is organized as follows. In Sec.\ref{II}, we introduce the physical settings for soliton transport firstly. Secondly, we present various soliton Thouless pumping scenarios with varying ISN, and demonstrate that soliton pumping can be engineered well by the ISN. Thirdly, we discuss the impact of nonlinear instabilities on soliton transport, and show that there could be a proper window for observing soliton pumping with the lower limit of sweeping rate required by the instabilities and the upper limit given by the usual adiabatic conditions. In Sec.\ref{III}, we examine soliton transport associated with bifurcations from different bands. We show that nonlinear interactions between solitons can break their pumping behaviors. We summarize and discuss our results in Sec. \ref{IV}.

\section{varied soliton transport induced by inter-site nonlinearities}\label{II}
\begin{figure}[htbp]
    \centering
    \includegraphics[width=0.48\textwidth]{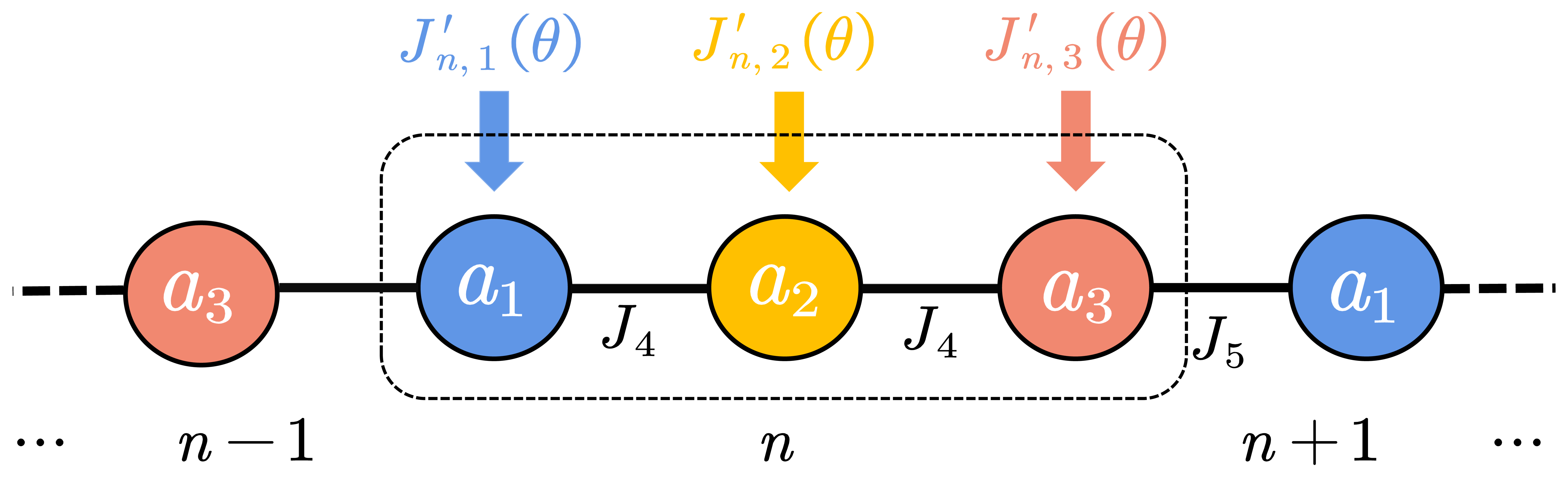}
    \caption{(a) Schematic of the diagonal extended AAH model. The dashed box labels a unit cell (indexed by $n$) containing three sites $(a_1, a_2, a_3)$, each subject to a modulated on-site potentials $J_{n,i}^{\prime}\left( \theta \right) =\sin \left[ \theta -\left( i-1 \right) 2\pi /3 \right] +\sum_j{\frac{g_{ij}}{2}\left| \psi _{n,i} \right|^2\left| \psi _{n,j} \right|^2}$ and $\psi_{n,i}$ denotes the wave-function at site $a_i$ in the $n$-th unit cell, with $\theta = \omega t$. Here, $J_4$ and $J_5$ represent the strength of linear hopping. $g_{ij} $ denotes the strength of nonlinear interactions.}
    \label{fig1}
\end{figure}
Most previous studies have focused on Kerr-type OSN ($g_{i\ne j}=0$), which depends solely on the local field intensity and typically does not break the correspondence between the linear CNs and soliton Thouless pumping. Motivated by the widespread presence of ISN in nonlinear systems \cite{DNLS,DNLS1} and the crucial role of inter-site couplings in determining topological properties in linear systems \cite{topological insulators1,topological insulators2,topological insulators3}, we expect that ISN may lead to soliton Thouless transport phenomena beyond the linear topology predictions. Thus we choose a one-dimensional AAH model with taking both OSN and ISN, shown in Fig. \ref{fig1}, to investigate soliton pumping behaviors. The dynamics of the system are governed by a discrete nonlinear Schr\"odinger equation \cite{DNLS, DNLS1, DNLS2}:
\begin{equation}
i\frac{\partial}{\partial t}\psi _{n,i}=\sum_{mj}{H_{ni ,mj}^{\mathrm{lin}}}\left( \theta \right) \psi _{m,j}+\sum_{j}{g_{ij}\left| \psi _{n,j} \right|^2\psi _{n,i}}
\label{NLS equation}
\end{equation}
where $\psi_{n,i}$ denotes the wave-function at site $a_i$ in the $n$-th unit cell. The subscripts $n,m$ label the unit cells along the one-dimensional chain, while $i,j = 1,2,3$ specify the three sites within each unit cell. The nonlinear coefficients $g_{ij}$ describe OSN for $i = j$ and ISN for $i\neq j$ with $g_{ij} = g_{ji}$. The first term in the evolution equation corresponds to the linear part of the system, whose Bloch Hamiltonian representation can be written as
\begin{equation}
H^{\mathrm{lin}}\left( k \right) =\left( \begin{matrix}
	J_1&		J_4\,\,&		J_5e^{-ik}\\
	J_4&		J_2&		J_4\,\,\\
	J_5e^{ik}&		J_4&		J_3\\
\end{matrix} \right).
\label{Bloch Hami}
\end{equation}
The on-site potentials are given by $J_{i=1,2,3} = \sin[\theta - (i-1)\, 2\pi/3]$, where $\theta(t) = \omega t$ is time dependent and $T = 2\pi / \omega $ denotes the driving period. The nearest-neighbor couplings are $J_4 = J_5 = 1$, describing the hopping between adjacent sites within a cell and between neighboring cells, respectively. The parameter $k$ denotes the quasimomentum, characterizing the Bloch states of the periodic system.
If the $\theta(t)$ is regarded as a synthetic dimension, the system forms a 2D three linear bands in the $(k, \theta(t))$ parameter space. In our case, the CNs of the three bands are $+1$, $-2$, and $+1$.
 In the presence of nonlinear interactions, soliton solutions exist at each instant of $\theta(t)$, which are also referred to as band-gap solitons \cite{gap soliton0, gap soliton1, gap soliton2}.

With setting $g_{ij} > 0$, the nonlinearity is de-focusing, and one family of instantaneous soliton solutions, bifurcating upward from the lowest energy band (with a CN of +1) is obtained using the Newton method while dynamically varying $\theta(t)$. When the adiabatic condition is satisfied, the soliton obtained from numerical evolution remains in excellent agreement with the corresponding instantaneous soliton solution throughout each full cycle.
Owing to probability conservation in a conservative quantum system, this family of soliton solutions typically requires fixing the particle number to be the same as that of the initial state. The soliton particle number, defined as $N_B = \sum_{n,i} |\psi_{ni}|^2$, remains conserved throughout the dynamical evolution. By the way, we find that it is more appropriate to define and fix the particle number only within the localized region in which the soliton resides when soliton radiation is present or more than one soliton cases are involved. Otherwise, the instantaneous soliton solutions will fail to describe the evolution of solitons obtained by numerical simulations.

\begin{figure}[htbp]
    \centering
    \includegraphics[width=0.5\textwidth]{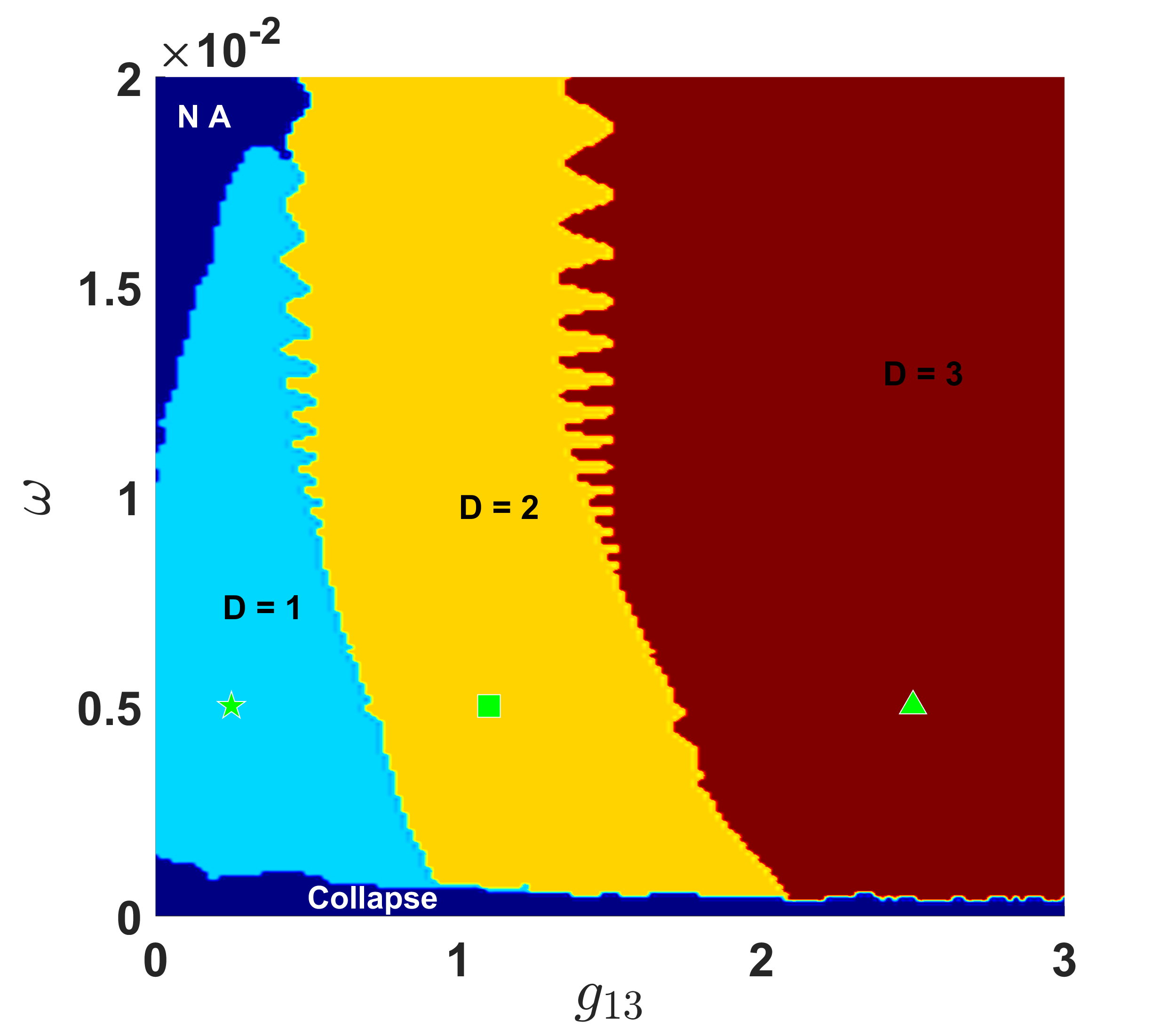}
    \caption{Phase diagram of the peak displacement (in units of cell) of solitons bifurcating from the lowest band over a cycle. From left to right, as $g_{13}$ increases, the solitons displacement per cycle exhibits quantized increments. From top to bottom, as $\omega$ decreases, the solitons fails to complete transport within a cycle, with the black numbers in the diagram indicating the displacement. In the upper-left region of the diagram, the solitons also fail to undergo transport as a result of non-adiabatic (NA) effects. Parameters: $n = 20, \mu = -1.1$. }
    \label{fig2}
\end{figure}

We choose solitons with the same chemical potential $\mu$ $(\mu =\frac{\langle \psi |H\left( \psi \right) |\psi \rangle}{N_B})$ at $\theta(t=0)$ to ensure that the initial solitons with different nonlinear interaction strengths are located at the same position within the band gap of the linear energy spectrum. By selecting different values for the ISN $g_{13}$ (unless otherwise specified, $g_{11}=g_{22}=g_{33}=g_{12}=1$) and driving frequencies $\omega$, the system is numerically evolved using the Runge-Kutta method \cite{ODERK 1980}.
The absolute magnitude of the ISN is irrelevant, only its strength relative to the OSN matters. In experiments, the ISN can be made comparable to or exceed the OSN, for instance, using Feshbach resonance \cite{Feshbach 2010} on a Bose-Einstein condensate in an optical lattice.
After one cycle, the system returns to its initial position, and the soliton, after undergoing a series of deformations, also recovers to its initial state. The fidelity between the initial and final soliton states is then calculated, which is used to determine whether transport has occurred. If the fidelity exceeds $96\%$, the soliton is regarded as having successfully completed the transport. The transport can equivalently be characterized by the soliton profile itself, namely, by the displacement of the soliton peak corresponding to the quantized transport distance. Conversely, if the fidelity falls below $96\%$, the soliton is taken to have failed to completed the transport. Correspondingly, a phase diagram of the soliton's peak displacement bifurcating from the lowest linear energy band can be constructed, as shown in Fig. \ref{fig2}. Notably, in the presence of self-crossing structures during soliton transport\cite{Gong Jiangbin 2023}, fidelity alone may not reliably characterize the transport, and such regions are therefore excluded from our analysis.

In Fig. \ref{fig2}, the light blue, yellow, and red regions denote the parameter ranges where the solitons successfully complete transport within one cycle, with $D$ representing the transport displacement. Interestingly, the soliton transport distance exhibits quantized increments as $g_{13}$ is increased, advancing from one unit cell to three unit cells, in sharp contrast to the only one unit cell predicted by topology of linear band. This clearly suggests that nonlinearities indeed bring topological effects beyond linear band topology.  There are oscillations boundaries between regions corresponding to different integer transport, which may indicate the presence of phase transitions.
Two dark blue regions in the phase diagram correspond to parameter ranges where the soliton fails to complete transport within a single cycle. In the upper-left region, the soliton transport is inhibited at relatively high driving frequencies $\omega$ due to non-adiabatic effects. Strikingly, at low driving frequencies limit, the soliton also fails to complete transport, seems to challenge the common notion that slower modulation enables a more favorable realization of topological transport. We uncover that the lower limits on sweeping rates are required by nonlinear instabilities, see details in the subsection B.

We emphasize that the anomalous soliton transport distance does not result from a combination of CNs associated with occupation of multiple Bloch bands \cite{Ye Fangwei 2022}. This can be checked by calculating the populations of soliton states on instantaneous linear Bloch states of each band.
The population of the soliton wave function $\psi _{n,i}^{s} \left( \theta \right)$ on the instantaneous linear Bloch states at any given time $\theta(t)$ can be obtained by:
\begin{equation}
\rho _l\left( \theta \right) =\frac{1}{N}\sum_k{\left| \sum_{n,i}{\psi _{l,n,i}^{\text{Bloch}}\left( k,\theta \right) ^*\psi _{n,i}^{s}\left( \theta \right)} \right|}^2
\label{Proportion}
\end{equation}
$\rho_l(\theta)$ denotes the occupation of the $l$-th band at time $\theta(t)$.
$\psi^{\mathrm{Bloch}}_{l,n,i}(k,\theta)^*$ denotes the complex conjugate of the linear Bloch state in the $l$-th band labeled by the quasimomentum $k$. The factor $1/N$ ensures normalization.
The occupation numbers $\rho_l(\theta)$ on the lowest energy band \(l = 1\) satisfy $\rho_{1,\mathrm{star}}(\theta) > 0.979$, $\rho_{1,\mathrm{rectangle}}(\theta) > 0.982$, and $\rho_{1,\mathrm{triangle}}(\theta) > 0.987$ within one period for solitons on the star, rectangle, and triangle in Fig. \ref{fig2}.
These indicate that the soliton states remain almost entirely confined to their corresponding bifurcated linear bands within one driving cycle. Namely, the topology of the linear bands remains unchanged and the soliton continues to occupy only a single linear band. But the CN of the lowest linear band fails to characterize transport involving more than two unit cells. Thus the extra pumping effects are purely induced by the nonlinearities in origin.

\subsection{Solitons pumping engineered by nonlinearities}

\begin{figure}[htbp]
    \centering
    \includegraphics[width=0.5\textwidth]{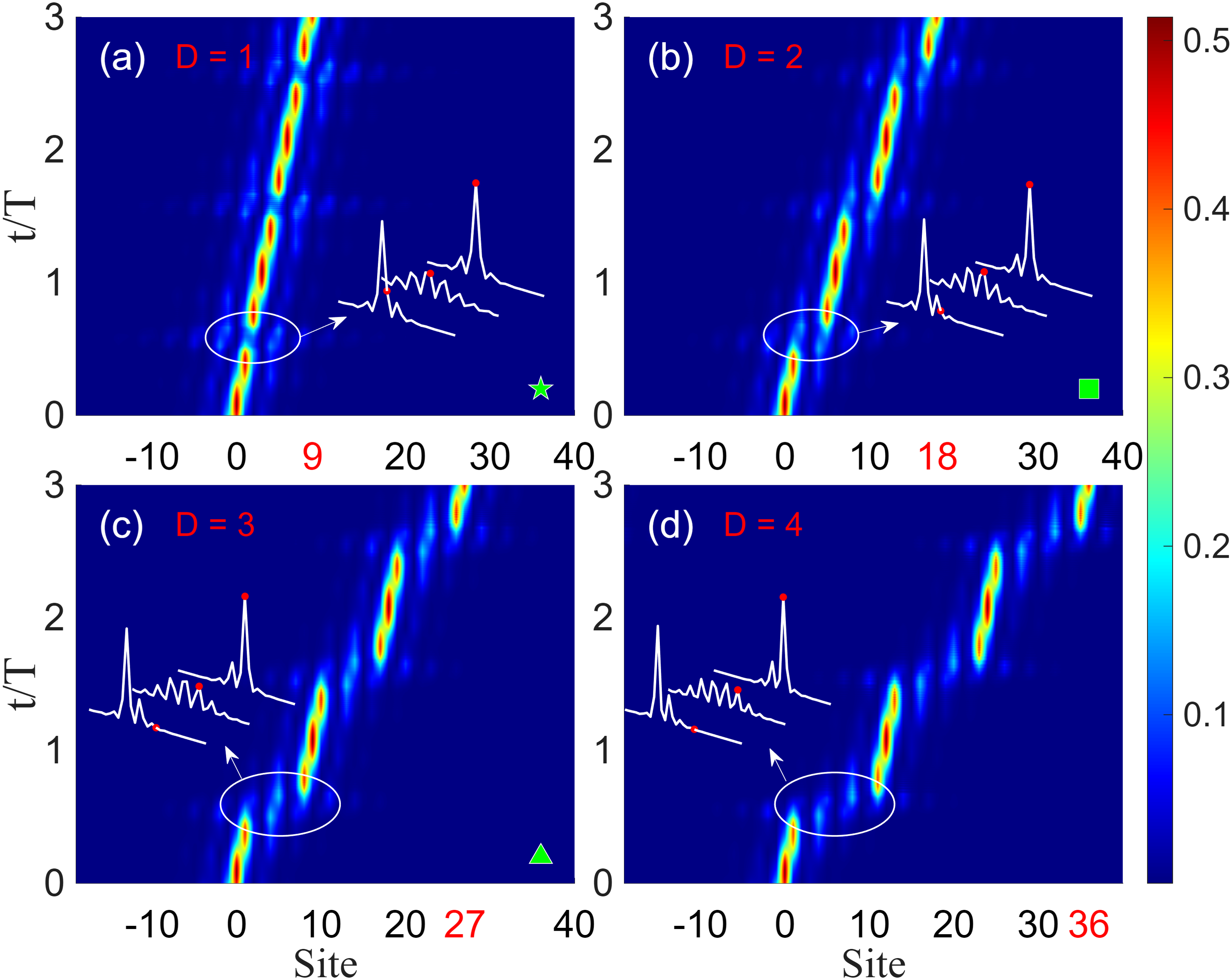}
    \caption{Evolution of varied solitons over three cycles. The red tick marks indicate the soliton peak positions after three cycles. In the inset, the white solid lines represent the soliton density distribution within the elliptical frame, with red dots corresponding to the same lattice site. (a) Normal transport: the soliton displacement per cycle exactly matches the CN of the bifurcating linear band, with $g_{13} = 0.25$. (b-d) Varied transport: the soliton displacement per cycle corresponds to 2, 3, and 4 unit cells. The marker shown in the lower-right corner represents the same parameters as indicated in Fig.\ref{fig2}. The parameters $ g_{13} $ are 1.1, 3, and 6.5, respectively. Common parameters: $ n = 20 $, $ \mu = -1.1 $, and $\omega = 0.005$.}
    \label{fig3}
\end{figure}

The close relations between soliton transport per cycle and values of $ g_{13} $ provide possible ways to engineer soliton pumping by nonlinearities. For examples, we show the evolution of solitons with $g_{13}= 0.25, 1.1, 3$ and $6.5$ in Fig. \ref{fig3}, which admit $D=1, 2, 3$ and $4$ per cycle, respectively. Very recently, it was demonstrated that nonlinearities can make soliton transport reach distances of up to twice the linear CN \cite{Xu Yong2 2025}. The ISN was found to suppress the anomalous transport, which is completely contrary to the results reported here. The soliton becomes broadened greatly during each pumping cycle, as illustrated in Fig. \ref{fig3}. Our analysis indicates that it happens when the time-dependent on-site potential reaches certain symmetric points, e.g. $\theta = \frac{(6m+1)\pi}{6}$, $m=1,3,5,\cdots $.  The solitons move almost in identical trajectory before the great broadened behavior emerging (marked by white circle), and they admit quite distinctive drifts after the broadened behavior. Thus the broadened behaviors play crucial roles in the varied topological pumping. We show soliton profiles around the broadened moment in the insets of Fig. \ref{fig3}.  It is seen that the solitons exhibit multi-peak profiles. The broadened solitons allows neighboring sub-peaks to reach amplitudes comparable to that of the main peak. Different $g_{13}$ settings can cause distinctive sub-peaks grow rapidly to be new main peaks, and larger $g_{13}$ values tend to induce more drift of solitons.  The main-peak and sub-peaks are localized at equivalent lattice sites, which makes drift be also quantized in unit of cells.

Although such diverse forms of soliton transport cannot be predicted by the linear CN, they can be described a posteriori within a supercell framework \cite{Jurgensen 2021,Xu Yong2 2025}. The nonlinearities inducing topology is still hidden, because the supercell method effectively treats the soliton as a non-periodic defect. Another possible viewpoints attribute this behavior to additional soliton solutions arising from the superposition of Wannier functions on neighboring sites \cite{Xu Yong2 2025}. A unified topology theoretical framework is still lacking for the pumping beyond linear band theory. The above mentioned multi-peak profile of solitons involve multiply nodes, which are inherited from the higher-order Bloch modes \cite{gap soliton3, gap soliton4}. The striking topology could be related with the nodal structures of wave functions \cite{Zhao 2025, Dirac 1931}.

\subsection{Instability of solitons can be inhibited by driving frequencies}

\begin{figure}[http]
    \centering
    \includegraphics[width=0.5\textwidth]{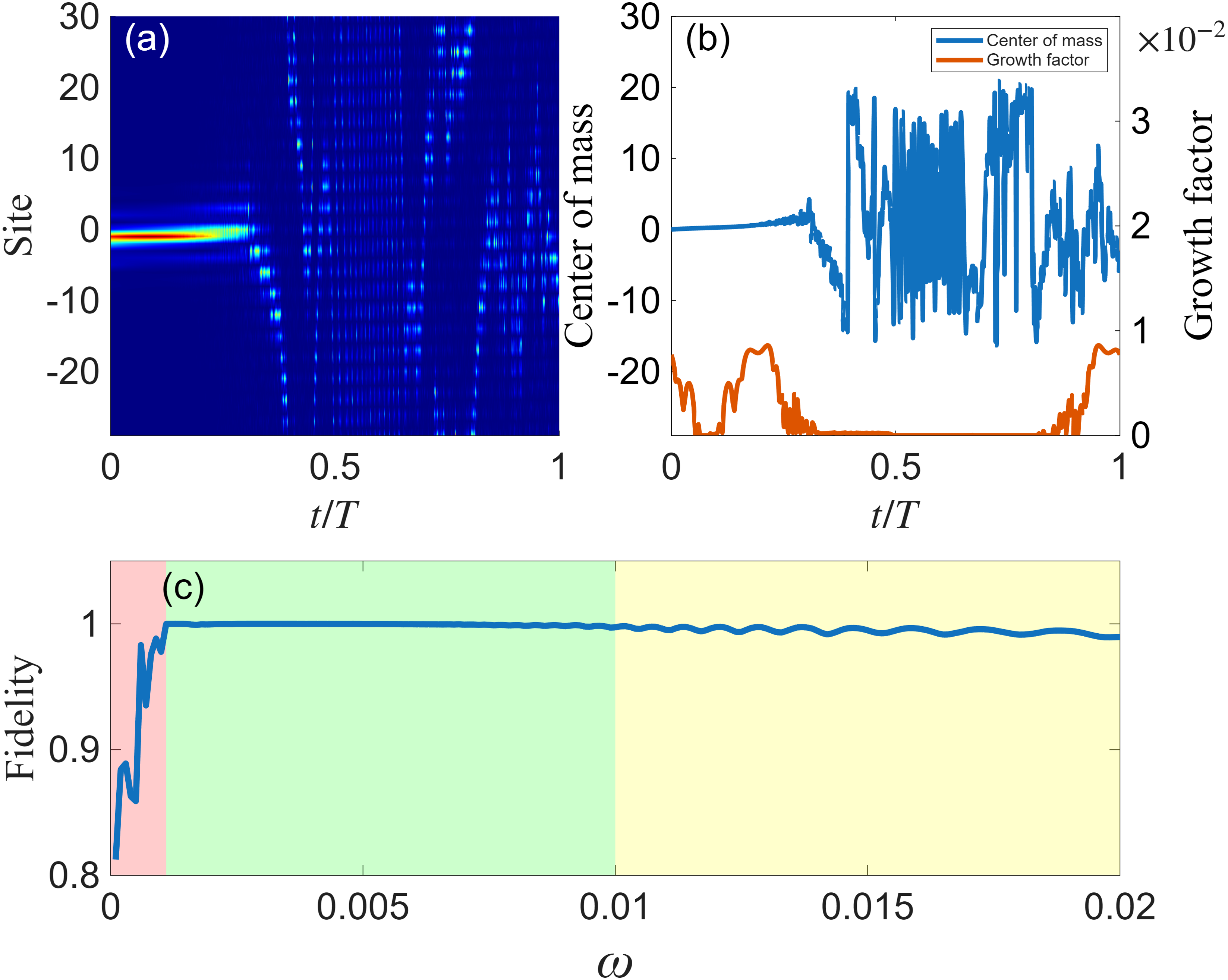}
    \caption{(a) Soliton destruction induced by weak instability. (b) Left: COM motion after soliton breakdown; right: linear stability analysis of the soliton within one cycle, showing the maximum growth rate at each instant, with $\omega = 0.0002$. (c) Soliton transport over one cycle, with the fidelity between the initial and final states shown for different driving frequencies. The fidelity fluctuations in the light yellow region on the right arise from NA, whereas the light red region on the left corresponds to soliton destruction caused by instability. Common parameters: $n = 20$, $g_{13} = 1.1$, and $\mu = -1.1$.}
    \label{fig4}
\end{figure}
\begin{figure}[http]
    \centering
    \includegraphics[width=0.5\textwidth]{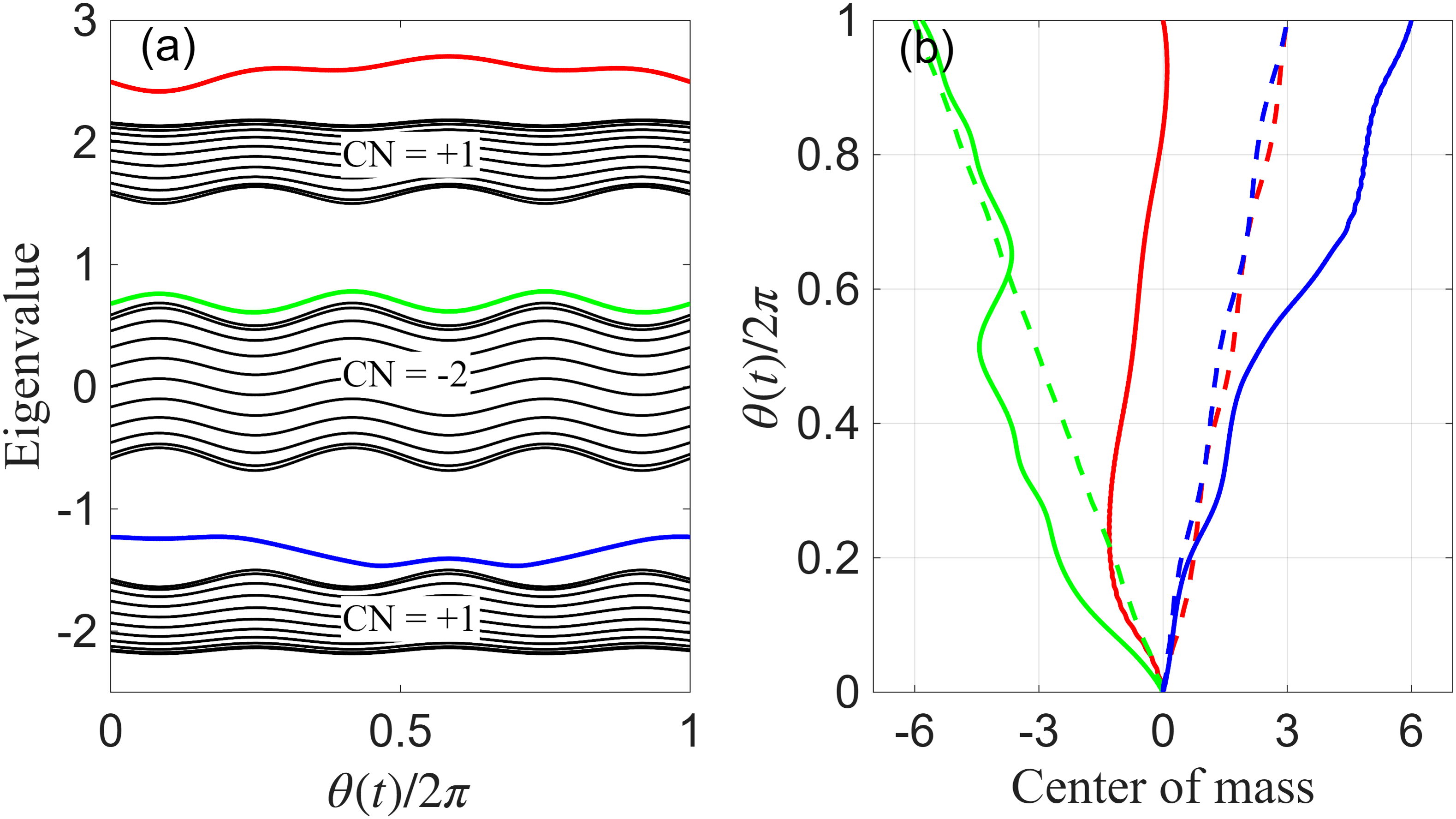}
    \caption{(a) Colored lines represent gap solitons bifurcating from different linear energy bands, all with the same particle number $N_B = 0.95$. The black regions indicate the linear energy bands. (b) Colored solid lines show the COM transport of solitons (colors correspond to those in (a) under the same driving frequency $\omega = 0.005$. The dashed lines of the same color indicate the corresponding COM transport of the Wannier states. Parameter : $g_{ij}=1, i,j = 1,2,3$}
    \label{fig5}
\end{figure}

Due to the presence of de-focusing nonlinear interactions, the solitons exhibiting anomalous transport always bifurcate upward from the lowest energy band and remain located within a finite band gap, as shown by the blue line in Fig. \ref{fig5}(a). In general, solitons residing in finite band gaps usually admit instability \cite{gap soliton3,gap solitons2 2008,gap solitons3 2003}. This makes soliton transport require a minimum driving frequency  (see Fig. \ref{fig2}), below which a complete transport cycle cannot be achieved. For an example, we show the evolution of soliton with $\omega=0.0002$ in Fig. \ref{fig4}(a). It is seen that the soliton break down and the pumping seems to be broken. The quantized soliton's center of mass (COM) transport is also absent, as shown by the blue solid line in Fig. \ref{fig4}(b)). Recent studies have shown that self-crossings of nonlinear bands can also disrupt the COM transport of solitons, even though the soliton peaks themselves continue to undergo transport \cite{Gong Jiangbin 2023}. In this work, the nonlinear bands do not exhibit such self-crossing behavior, as illustrated by the colored solid lines in Fig. \ref{fig5}(a).

To quantify their instability, we perform a linear stability analysis of a sequence of instantaneous intrinsic soliton solutions within a single driving cycle, and extract the largest growth rate at each instant. As illustrated by the orange solid line in Fig. \ref{fig4}(b),  soliton indeed exhibits weak instability during each cycle.  Such instabilities could generally disrupt the topological transport of solitons. Fortunately, the instabilities destructive effect takes a certain amount of time to accumulate before it can manifest itself, and the magnitude of instabilities depend on the soliton forms which is varying with evolution. Especially, there are some stability regimes during each driving cycle, see the orange solid line in Fig. \ref{fig4}(b). These characters provide possibilities to inhibit them by increasing the driving frequency. For example, multi-cycle soliton transport can be achieved with $\omega = 0.005$, as illustrated in Fig. \ref{fig3}(b), in contrast to the ones in Fig. \ref{fig4}(a) with identical other parameters settings. There is a competition between soliton instability and adiabaticity, which jointly require the conditions of soliton transport. Consequently, the pumping conditions are restricted to an intermediate window of driving frequencies (the green region in Fig. \ref{fig4}(c)), bounded from below by soliton instability (the red region) and from above by the breakdown of adiabaticity (the yellow region). In the red region, the soliton breaks down quickly, and this behavior can  be denoted by the fidelity between initial state and final state of a full cycle, which is much lower than one. In the green region, the fidelity is one almost, which indicates the final soliton return to initial form perfectly and multi-cycle pumping can be observed. In the yellow region, the fidelity deviates from one slowly, and non-adiabatic effects begin to affect the pumping process.
The weakly unstable on-site solitons can still be used to realize topological pumping with proper driving frequencies, in contrast to
the  inter-site gap solitons  which typically exhibit strong instabilities and rapidly collapse during soliton transport \cite{Jurgensen 2022}.

\section{Distinct transport behaviors of soliton energy levels bifurcating from different linear energy bands}\label{III}

\subsection{Thouless pumping of solitons bifurcating from different linear energy bands}

It is well known that solitons bifurcating from different linear energy bands should admit distinctive bifurcation direction, waveform structure and stability \cite{Solitons 2011}. We further check whether the ISN affects the transport of solitons bifurcating from different linear energy bands in distinctive ways, while the pumping behaviors beyond linear band topology.  With fixing the nonlinear coefficient $g_{i,j} = 1$ and the particle number $N_B = 0.95$, a family of soliton solutions bifurcates upward from each Bloch band (see the colored solid lines in Fig. \ref{fig5}(a)).
The energy levels are obtained via the Newton method from the initial state $\psi_{try}=A \sech(\eta x) \psi_{l}^{\mathrm{Bloch}}(k,0)$, where $\psi_{l}^{\mathrm{Bloch}}(k,0)$ is the Bloch state of one linear band.  $k = 0$ or $\pi$ is chosen to select the Bloch state at either the lower or upper edge of each linear band \cite{Solitons 2011,gap soliton3}.  $A \sech(\eta x)$ denotes a soliton envelope determining particle number, where $A$ and $\eta$ control the amplitude and width of the soliton. Solving the nonlinear eigenvalue problem with the fixed particle number of soliton, we obtain the three eigenvalues with choosing the Bloch states of three different linear bands separately. The energy levels are reasonably regarded as soliton energy levels, since the eigenstates correspond  soliton wavefunctions.  The calculated soliton energy levels indeed describe the soliton transport behavior, in the sense that the corresponding eigenstates are consistent with the soliton states during the pumping process. We calculated the transport trajectories of the solitons COM with the driving frequency $\omega = 0.005$, shown by solid lines in Fig. \ref{fig5}(b).
To see the differences in comparison with linear band topology theory, we calculate the corresponding Wannier states COM of the bifurcating bands with setting $g_{\alpha,\beta} = 0$ and keeping other parameters,  shown by the dashed lines in Fig. \ref{fig5}(b). Strikingly, the transport displacements of solitons with lower, middle, and upper energy levels admit multiple (beyond CN of the linear band), normal (identical with CN of the linear band), and trapping (no pumping), respectively.  Even though solitons from the middle band undergo normal transport, their COM trajectories differ significantly from those of the corresponding Wannier states.
By the way, it should be noted that the trapping is induced by pure nonlinearities, also departing from the CN of its occupying band. The calculation based on Eq. \ref{Proportion} shows that each soliton remains predominantly confined to its bifurcating linear band during the evolution, there are no other bands involved for each soliton. This trapping is quite different from trapping phenomena involving multi-bands in previous studies \cite{Ye Fangwei 2022}.

The upward bifurcation of soliton will become downward bifurcation for the focusing nonlinearities $g_{ij} < 0$. In this case,  the solitons bifurcating from the lowest and highest bands admit trapping and multiple pumping separately, which are inverse in comparison with the above ones with de-focusing cases. Interestingly, our simulation results suggest that the trapping of the soliton bifurcating from another band always emerge when there are  multiple pumping for soliton with one band. This hints that there are some possible constrains on the pumping behaviors of solitons in our model with ISN.

\begin{figure}[http]
    \centering
    \includegraphics[width=0.5\textwidth]{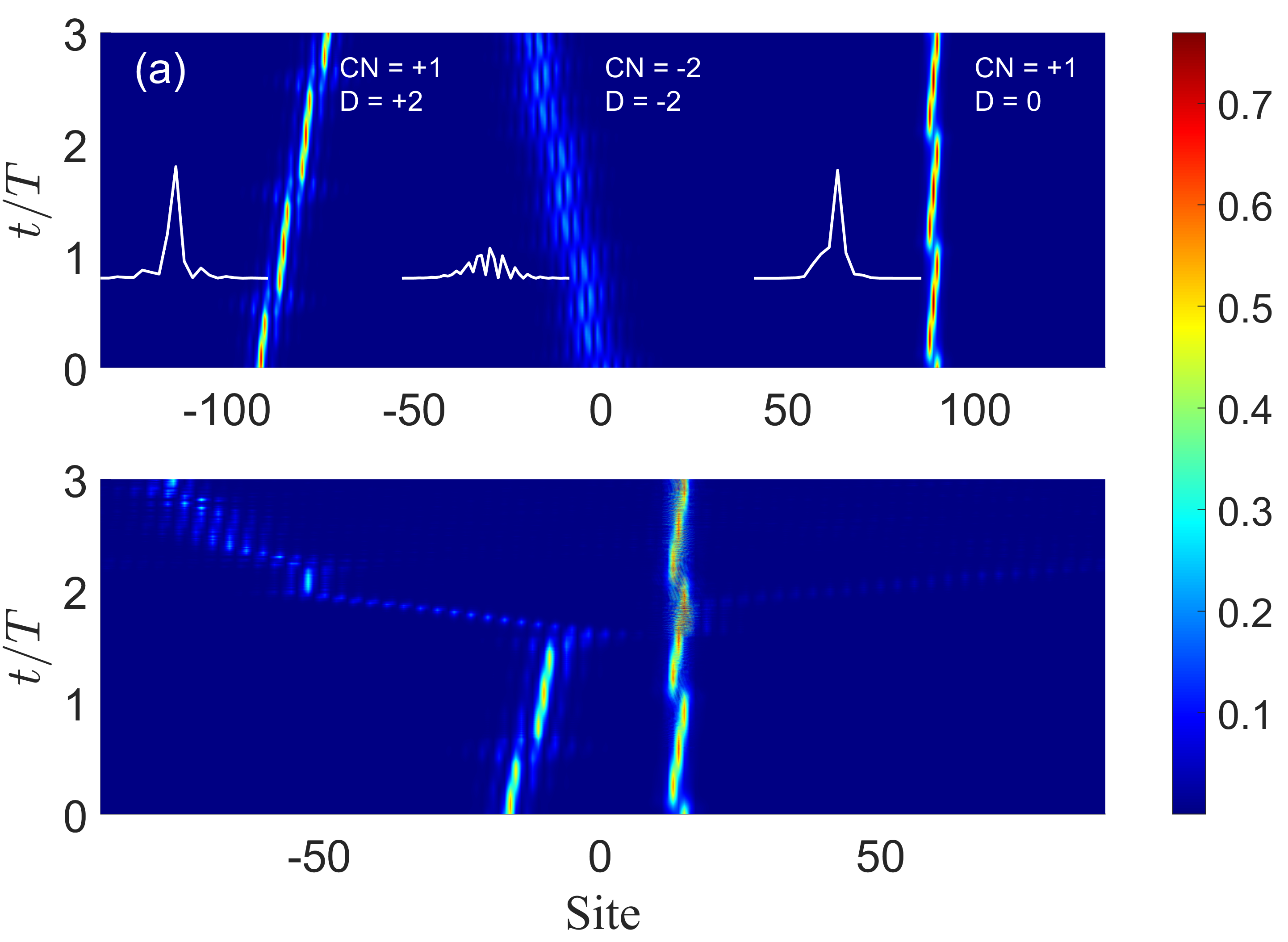}
    \caption{Multi-soliton transport. (a) From left to right: transport of solitons bifurcating from the lower, middle, and upper bands (as indicated in Fig.\ref{fig5}); the solitons are well separated. The insets show the soliton profiles at $t=0$. (b) Transport of solitons bifurcating from the lower and upper bands; the interactions between solitons can not be ignorable. Parameters: $n=90$, $\omega=0.005$, $N_B = 0.95$, $g_{ij}=1, i,j = 1,2,3$.}
    \label{fig6}
\end{figure}

\subsection{Thouless pumping of multi-solitons}

One can construct several solitons in which each individual soliton bifurcates from a linear energy band, and the linear band can be chosen freely even if soliton energy level exists. This allows us to investigate multi-soliton transport cases. The total chemical potential $\mu$ of the system, given by the nonlinear eigen-equations with total particle number of multi-soliton, no longer possesses a clear physical meaning, since its value no longer provides information about the individual solitons. We propose that the soliton energy levels should be calculated separately based on nonlinear eigen-equations with calculating particle number of each soliton within its localized region. When the solitons are well separated, they can be regarded as effectively independent during transport, as shown in Fig. \ref{fig6}(a). Only the pumping behavior of the middle soliton corresponds to the calculated CN. Moreover, the calculation of soliton energy levels should be performed based on the particle number of the localized soliton wave packet, rather than the total particle number of the system, especially when radiation occurs during soliton transport \cite{Gong Jiangbin 2023}. However, nonlinear interactions between solitons can vary their pumping behaviors when there are obvious overlapping between solitons. As an example, we choose the left one and right one in Fig. \ref{fig6}(a) to demonstrate how nonlinear interactions vary solitons pumping. The restuls are shown in Fig. \ref{fig6}(b). This demonstrates that the left one firstly can transport with a quantized pumping, but the soliton is destroyed totally when it interacts with the other soliton.  The right soliton still admits pumping well only with some noises and weak rediations. Thus soliton pumping generally lacks the robustness characteristic of Thouless pumping as linear systems. In linear systems, the transport of particles or wave packets is mutually independent. In nonlinear systems, collisions between solitons can essentially affect their pumping and could provide alternative ways to manipulating pumping dynamics.

\section{conclusion and discussion}\label{IV}

We show that ISN can be used to engineer soliton pumping in a diagonal AAH model, for which solitons can possess displacements of 2, 3, or even 4 unit cells  within a single pumping cycle, which far exceeding the CN$=+1$ of the corresponding linear band. The anomalous extra transport of solitons is purely induced by nonlinearities, and the soliton states do not occupy multiple linear bands during their evolution, in contrast to pumping processes involving multiple linear bands  \cite{Ye Fangwei 2022}.  We find that instabilities bring clear lower limits on sweeping rates. We further investigate pumping of multi-solitons. When the solitons are well separated, they can be regarded as effectively independent during transport. Soliton energy levels should be calculated separately, and the results are in excellent agreement with the pumping behavior of well-separated solitons. However, nonlinear interactions between solitons can vary their pumping behaviors greatly when there are obvious overlapping between solitons.

Complex factors such as non-Hermiticity and disorder can be introduced to investigate their synergistic effects with nonlinearities on topological transport \cite{non-Hermitian 2021}, thereby enhancing the relevance between theory and experiments. The nonlinear interactions among multi-solitons can not only undermine transport robustness but also provide new degrees of freedom for topological regulation. In the future, research can focus on realizing controllable switching of topological phases through soliton collisions and overlaps, and developing topological devices \cite{photonics 2019} such as topological switches based on multi-soliton interactions. Meanwhile, exploring topological protection mechanisms in multi-soliton systems and designing multi-channel topological transport systems with anti-interference capability are promising directions.

\section*{Acknowledgments}
 This work is supported by the National Natural Science Foundation of China (Contract No. 12375005, 12235007, 12247103).

\end{document}